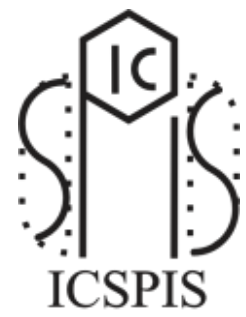

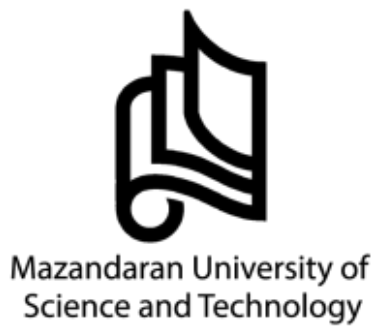

# A Hybrid Cluster-Based Classification Model for Anomaly Detection in Unbalanced IoT Networks

Hossein Shaemi Barzoki
Faculty of Computer Engineering
University of Isfahan
Isfahan, Iran
hossein.shaemi@mehr.ui.ac.ir

Amir Hossein Fathi Hafshejani
Faculty of Computer Engineering
University of Isfahan
Isfahan, Iran
fathiamir@eng.ui.ac.ir

Ahmadreza Montazerolghaem
Faculty of Computer Engineering
University of Isfahan
Isfahan, Iran
a.montazerolghaem@comp.ui.ac.ir

***Abstract*— Detecting anomalies in Internet of Things (IoT) networks is a critical security challenge, often hampered by highly imbalanced and diverse network traffic datasets. Standard classifiers struggle to perform well across all traffic types. This paper proposes a hybrid detection model to address this challenge using the Bot-IoT dataset. Instead of a single complex classifier, we first employ K-Means clustering to segment the training data into three distinct traffic profile clusters. We then train and evaluate multiple baseline machine learning models, including Decision Tree, KNN, and XGBoost, on each cluster independently to identify the optimal classifier for that specific data profile. Our results show that this cluster-specific, hybrid approach, which assigns different simple models to different clusters, improves detection accuracy and provides a more robust and efficient framework for handling diverse IoT attack traffic.**

***Keywords— Internet of Things (IoT), Anomaly Detection, Intrusion Detection, Bot-IoT, Imbalanced Data, K-Means Clustering, Machine Learning***

## I. INTRODUCTION (*HEADING 1*)

With the rapid development of communication technology, IoT is growing at an explosive rate, becoming an essential part of industrial applications, medical services, and smart manufacturing [1],[2]. However, this explosive growth also introduces significant security challenges. The resource-constrained nature of many IoT devices, including low computational power and memory, makes it impractical to employ traditional, heavyweight security mechanisms[3],[4]. As a result, these networks are highly susceptible to attacks. Conventional, signature-based Intrusion Detection Systems (IDS) are often unable to repel the dynamic and novel threats targeting IoT devices [3], [4]. Consequently, data-driven methods using Machine Learning (ML) and Deep Learning (DL) have emerged as promising solutions for identifying sophisticated patterns and anomalies in massive network data [5], [6].

Despite their promise, these ML/DL models face two fundamental challenges when applied to real-world IoT traffic, such as that found in the Bot-IoT dataset. The first is severe class imbalance, where "normal" traffic is a tiny minority, causing models to develop a strong bias [10]. While many studies address this with data-level techniques like SMOTE [7], [10], these can introduce their own biases or lead to information loss.

The second, related challenge is data heterogeneity. A single "generalist" model must learn to distinguish "normal" behavior from a wide, diverse, and evolving range of attack types (e.g., DoS, DDoS, Reconnaissance, Theft). This complexity can force a model to find a sub-optimal, "one-size-fits-all" solution. Some research has already demonstrated the value of data partitioning, such as by splitting datasets into balanced parts to improve model training [9].

This paper proposes that these challenges are best solved not by a single, complex model, but by a hybrid "divide and conquer" strategy. We hypothesize that the heterogeneous Bot-IoT dataset is not one single problem, but rather a collection of smaller, more homogenous sub-problems.

We introduce a hybrid cluster-based framework. Instead of training one "generalist" classifier on the entire dataset, we first employ an unsupervised clustering algorithm (e.g., K-Means) to partition the training data into distinct traffic profiles. Each resulting cluster represents a more homogenous subset of the data based on feature similarity (e.g., clusters with distinct traffic patterns or attack characteristics), with each maintaining both normal and attack samples. We then train a separate, simpler, ML model on each cluster independently.

This approach allows us to deploy simple, highly-optimized models for each specific traffic profile, rather than one complex model struggling to learn all patterns at once. We will demonstrate that this hybrid, cluster-based framework provides a more robust and accurate method for anomaly detection, particularly in handling the severe class imbalance and heterogeneity of modern IoT networks.

While data-level approaches like resampling can mitigate class imbalance to some extent, they often operate under the assumption that the underlying data distribution is monolithic. This assumption is frequently violated in complex IoT networks, where attack vectors such as DDoS, reconnaissance, and data exfiltration exhibit fundamentally different characteristics from each other and from normal traffic. Applying a global resampling strategy to such a heterogeneous dataset can inadvertently amplify noise or create spurious patterns that do not generalize well, as the synthetic samples may not respect the distinct latent structures inherent to different attack types [9]. Therefore, treating the dataset as a single entity for balancing and modeling remains a suboptimal solution.

To address this fundamental limitation, we posit that a more effective strategy involves decomposing the problem space itself. By leveraging unsupervised learning to identify natural groupings within the data, we can transition from a single, complex classification task to multiple, simpler, and more homogeneous sub-tasks. This paradigm shift allows for the application of tailored solutions that are precisely calibrated to the specific characteristics of each data partition. Clustering serves as a powerful precursor to classification, effectively disentangling the intertwined distributions of various traffic profiles and paving the way for more accurate and specialized model training.

Consequently, this paper introduces a hybrid framework that systematically integrates clustering and classification. Our core contribution lies in a "divide and conquer" methodology where we first employ K-Means clustering to partition the preprocessed Bot-IoT dataset into distinct, homogeneous clusters. Subsequently, we train and select an ensemble of specialized machine learning models, each optimized for the unique profile of its assigned cluster. This approach not only alleviates the challenges posed by severe class imbalance and data heterogeneity but also demonstrates that a collection of simpler, interpretable models can collectively outperform a single, monolithic classifier in accuratelyidentifyingnetwork anomalies in IoT environments.

## II. Related Work

Numerous studies have explored machine learning and deep learning techniques for IoT intrusion detection. Al-Garadi et al. [5] provided a comprehensive survey of ML and DL methods applied to IoT security, highlighting the potential of these approaches but also noting challenges such as data imbalance andcomputational constraints. Bharati and Podder [6] further discussed the applications of ML/DL in IoT security, emphasizing the need for lightweight and efficient models.

To address class imbalance, resampling techniques like SMOTE [7] and RandomUnderSampler have been widely adopted. Bagui and Li [10] demonstrated the effectiveness of resampling methods in improving the performance of IDS on imbalanced datasets. However, these methods may introduce noise or overfit the minority class.

Another line of research focuses on data partitioning and ensemble methods. Atuhurra et al. [9] proposed splitting imbalanced datasets into balanced subsets to enhance model performance. Similarly, Zeeshan et al. [8] used protocol-based segmentation to improve detection accuracy for DoS and DDoS attacks. These studies underscore the value of decomposing complex data into manageable subsets.

Unlike previous work, our approach integrates unsupervised clustering with specialized model training. By first grouping data into homogeneous clusters and then training cluster-specific models, we address both imbalance and heterogeneity simultaneously. This hybrid framework builds on the strengths of both clusteringand classification, offering a scalable and efficient solution for IoT anomaly detection.

Furthermore, while ensemble methods like Random Forests and XGBoost have demonstrated robust performance in general classification tasks, their application to highly imbalanced and heterogeneous IoT security datasets often reveals inherent limitations. These models, by design, aim to build a global predictorthat optimizes performance across the entire data distribution. However, in scenarios where the feature space is partitioned into distinct regions representing vastly different attack signatures and normal behavior, a single complex ensemble may still struggle to capture the nuanced decision boundaries required for each specific traffic profile. This underscores the need for a more structural decomposition of the problem space beyond simply combining multiple models.

The concept of problem decomposition through clustering prior to classification has been explored in other domains with high data heterogeneity. However, its application to IoT intrusion detection, particularly with the Bot-IoT dataset, remains relatively nascent. The few studies that have ventured into this area often predefine clusters based on explicit, high-level features such as protocol type [8]. While beneficial, this manual partitioning may not capture the more subtle, multi-dimensional similarities between attacks that an unsupervised algorithm like K-Means can reveal. An automated, data-driven clustering approach applied to a comprehensive set of network flow features can potentially uncover more intrinsic and effective groupings for subsequent specialized modeling.

Our work directly builds upon and extends these ideas by proposing a tightly integrated, hybrid cluster-based classification framework. We move beyond using clustering merely for exploratory analysis or simple pre-filtering. Instead, we formally investigate the synergy between unsupervised clustering (K-Means) and a suite of supervised models, rigorously evaluatingthe optimal model selection for each identified cluster. This systematic approach to creating a ensemble of models, whereeachmodel is explicitlytailored to a homogenous data partition identified by clustering, addresses the dual challenges of imbalance and heterogeneity in a more principled manner than prior efforts. It represents a significant step towardsbuildingmore accurate, efficient, and interpretable intrusion detection systems for complex IoT ecosystems.

Recent studies have also begun to explore hybrid models that combine unsupervised and supervised learning paradigms. For instance, some researchers have utilized clustering as a pre-processing step to filter noise or identify broad patterns before applying a single classifier to the entire dataset. While these approaches acknowledge the value of understanding data structure, they often fail to fully leverage the potential of clustering to create specialized decision spaces. The transition from a clustered representation to a final classification is typically done in a loosely coupled manner, missing the opportunity to train dedicated models that are experts within each coherent data subspace. This indicates a gap for a more tightly integrated framework where clustering directly informs the creation of a specialized ensemble. Our work is distinguished from these predecessors by its holistic and systematic integration of clustering and classification into a unified, cluster-specialized ensemble framework. Unlike studies that use clustering merely for data explorationor preliminaryfiltering, we formally structure our entire modeling pipeline around the identified clusters. We rigorously evaluate a diverse set of simple, interpretable models to act as the selected model within each cluster, moving beyond the convention of employing a single complex model or a homogeneous ensemble. This approach ensures that the unique characteristics of each data partition—be it a specific type of Denial-of-Service attack or a cluster of normal operational traffic—are captured by a model specifically chosen and optimized for that context.

This represents a significant methodological advancement in constructing highly accurate and efficient intrusion detection systems tailored for the complex reality of IoT networks.

## III. Methodology

Our proposed method is a hybrid approach designed to handle the high imbalance and diverse traffic patterns within the Bot- IoT dataset. The methodology is executed in two main phases: 1) Data Preprocessingand Balancing, and 2) our proposed Hybrid Cluster-Based Classification Model.

Keep your text and graphic files separate until after the text has been formatted and styled. Do not use hard tabs, and limit use of hard returns to only one return at the end of a paragraph. Do not add any kind of pagination anywhere in the paper. Do not number text heads-the template will do that for you.

### *A. Data Preprocessing and Balancing*

The foundation of this study is the **Bot-IoT dataset**, a comprehensive andrealistic networktraffic dataset containing a wide variety of normal IoT traffic and cyber-attacks. A primary challenge of this dataset is its severe class imbalance, where attack records vastly outnumber normal records (approx. 99.9% to 0.013%). To prepare the data for modeling, we performed several key preprocessing steps :

- **Feature Reduction:** Non-essential identifier columns such as pkSeqID and seq were removed as they provide no predictive value for classification .
- **Data Transformation:** Port values (sport, dport) stored as hexadecimal strings (e.g., '0x...') were converted to their integer-based decimal equivalents
- **Categorical Encoding:** Text-based features like proto, saddr, and daddr were converted into numerical values using LabelEncoder to make them compatible with machine learning algorithms .
- **Data Balancing:** To address the severe class imbalance, we applied a combination of sampling techniques. We primarily used **SMOTE (Synthetic Minority Over- sampling Technique)** to generate synthetic samples for the minority class (normal traffic) and **RandomUnderSampler** to reduce thesize of the majority class (attack traffic) . This created a more balanceddataset for model. The comparative accuracy of models based on different resampling strategies is detailed in **Table 1**, which informed our final approach.

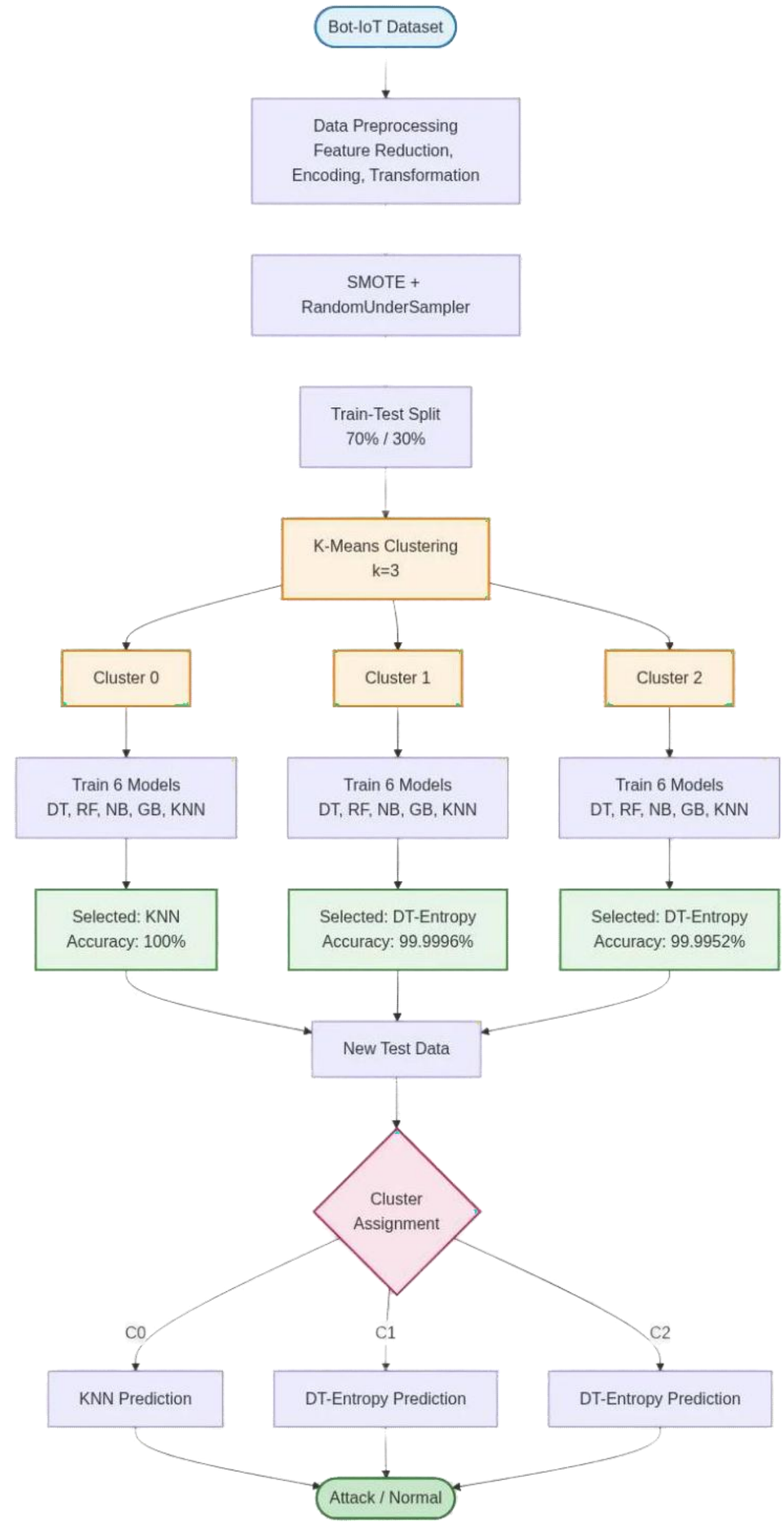

*Figure 1 Overview of the Hybrid Cluster-Based Classification Framework*

### *B. Proposed Hybrid Cluster-Based Model*

- Instead of relying on a single, complex classifier for the entire dataset, we developed a model framework. This framework, detailed in Figure 1, operates as follows:
- Step 1: Data Clustering: We first apply the K- Means clustering algorithm to the balanced training dataset.

Table 1. Comparative Accuracy of Models across three resampling strategies

| Model | **Cluster** | **Approach 1 (No Resample)** | **Approach 2 (Resample each cluster)** | **Approach 3 (Resample all data)** |
|---|---|---|---|---|
| **dtGini** | 0 | 1.0 | 0.999995963 | **1.0** |
| **dtGini** | 1 | 0.999996796 | 0.999996796 | 0.999930753 |
| **dtGini** | 2 | 0.999933481 | 0.999942984 | **1.0** |
| **dtEntropy** | 0 | 0.999995963 | 0.999995963 | 0.999998398 |
| **dtEntropy** | 1 | 0.999996796 | 0.999996796 | 0.999997230 |
| **dtEntropy** | 2 | 0.999947735 | 0.999952486 | **1.0** |
| **rf** | 0 | 1.0 | 1.0 | 0.999983980 |
| **rf** | 1 | 0.999871877 | 0.999996796 | 0.994396603 |
| **rf** | 2 | 0.996674078 | 0.985917099 | **1.0** |
| **nb** | 0 | 0.999991926 | 0.999995963 | 0.999288733 |
| **nb** | 1 | 0.999996796 | 0.999990390 | 0.918914165 |
| **nb** | 2 | 0.989718151 | 0.977854115 | **0.999897054** |
| **gb** | 0 | 1.0 | 1.0 | **1.0** |
| **gb** | 1 | 0.999839846 | 0.999903907 | 0.999958452 |
| **gb** | 2 | 0.999928730 | 0.999966740 | **1.0** |
| **knn** | 0 | 0.999995963 | 1.0 | **1.0** |
| **knn** | 1 | 0.999996796 | 0.999996796 | 0.999990305 |
| **knn** | 2 | 0.999938232 | 0.999938232 | 0.999948527 |

- Using the "Elbow Method" to determine the optimal number of clusters, we segmented the data into three distinct clusters (k=3). It is important to note that each cluster maintains the binary classification structure, containing both normal and attack traffic samples. The clustering does not separate classes but rather groups samples based on feature similarity, creating three distinct traffic profiles with different underlying patterns. This allows us to train specialized binary classifiers optimized for each profile's unique characteristics.
  - Step 2: Independent Model Training: We then train a suite of baseline machine learning models—including Decision Tree (gini and entropy), Random Forest, Gaussian Naive Bayes, XGBoost, and K-Neighbors Classifier (KNN)—on each of the three clusters independently.
  - Step 3: Model Selection: By evaluating the accuracy of each model within its cluster, we selected the single best-performing model for each cluster. Our findings identified KNN as the optimal model for Cluster 0 (see confusion matrix in **Fig. 2**), and Decision Tree (entropy) as the optimal model for both Cluster 1 (**Fig. 3**) and Cluster 2 (**Fig. 4**).
  - Step 4: Final Prediction Framework: The fully trained framework classifies new, unseen test data by first assigning it to the nearest cluster (0, 1, or 2). Then, only the dedicated model for that cluster (e.g., KNN for Cluster 0) is used to make the final prediction of "Attack" or "Normal".

This hybrid approach allows for the use of simpler, more efficient models that are highly optimized for a specific subset of the data, rather than a single generalist model that must account for all traffic variations.

### *C.* Model Evaluation and Validation Strategy

To ensure the robustness and generalizability of our hybrid framework, we implemented a rigorous evaluation strategy. Each specialist model's performance was evaluated primarily using accuracy as the selection metric. Given that the resampling strategy (SMOTE + RandomUnderSampler) effectively balanced the classes within each cluster, accuracy provides a reliable indicator of model performance. The entire process—from clustering to final prediction—was validated using a strict train-test split, where the test set was entirely held out from the initial clustering and balancing phases.

## IV. Using the Template

### *A. Experimental Setup*

The preprocessed and balanced data was split into training and testing sets, with 70% used for training the clustering and classification models and 30% held out for final validation. The six baseline models evaluated on each cluster were: **Decision Tree (gini)**, **Decision Tree (entropy)**, **Random Forest**, **Gaussian Naive Bayes**, **XGBoost**, and **K-Neighbors Classifier (KNN)** . Accuracy was used as the primary metric for selecting the best model for each cluster.

### *B. Model Selection*

The core of our experiment was determining the best model for each of the three data clusters. After training all six models on the data from each cluster, we found that the optimal model varied significantly, as shown in Table 2.

Table 2. Results of models on clusters

| Model | Cluster 0 Accuracy | Cluster 1 Accuracy | Cluster 2 Accuracy |
|---|---|---|---|
| dtGini | 0.999995 | 0.999996 | 0.999942 |
| dtEntropy | 0.999995 | 0.999996 | 0.999952 |
| rf | **1.0** | 0.999996 | 0.985917 |
| nb | 0.999995 | 0.999990 | 0.977854 |
| gb | **1.0** | 0.999983 | 0.999966 |
| knn | **1.0** | 0.999996 | 0.999938 |

The results clearly indicate that a one-size-fits-all model is not the most effective approach.

## V. Conclusion

This research addressed the significant challenge of anomaly detection in the severely imbalanced and heterogeneous Bot-IoT dataset. We demonstrated that a hybrid, cluster-based framework provides a highly accurate and efficient solution, moving beyond a one-size-fits-all classification approach. By segmenting the heterogeneous data into three homogenous clusters using K-Means, we were able to train simple, baseline models (specifically KNN and Decision Tree) that were optimized for each data profile. This methodology of assigning the best model to each cluster resulted in near-perfect accuracy on the unseen test data, proving its effectiveness and validating our hypothesis. The results show that this hybrid framework is a highly effective and robust strategy for anomaly detection in diverse IoT environments. For future work, this framework could be enhanced. We suggest exploring advanced synthetic data generation techniques, such as Generative Adversarial Networks (GANs), to further improve the quality of the minority class data. Furthermore, developing this framework into an adaptive, real-time system that can update clusters and models as new threats emerge would be a valuable next step in securing IoT systems.

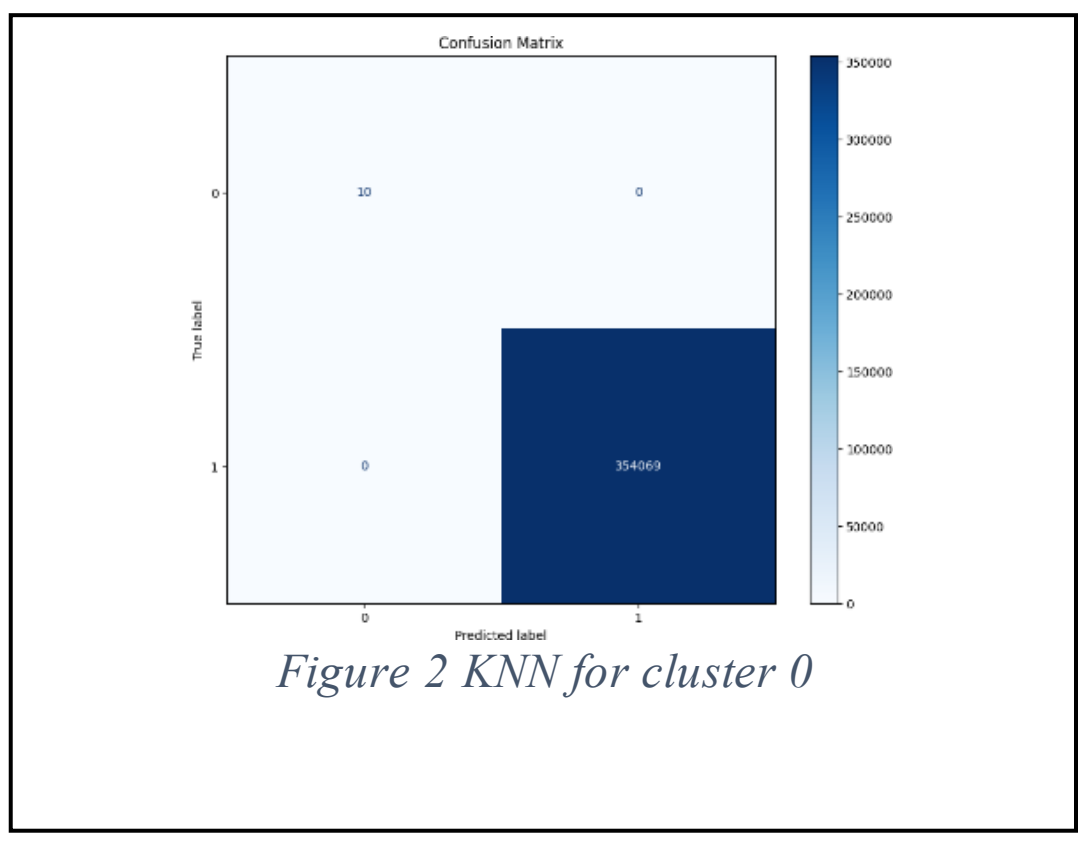

*Figure 2 KNN for cluster 0*

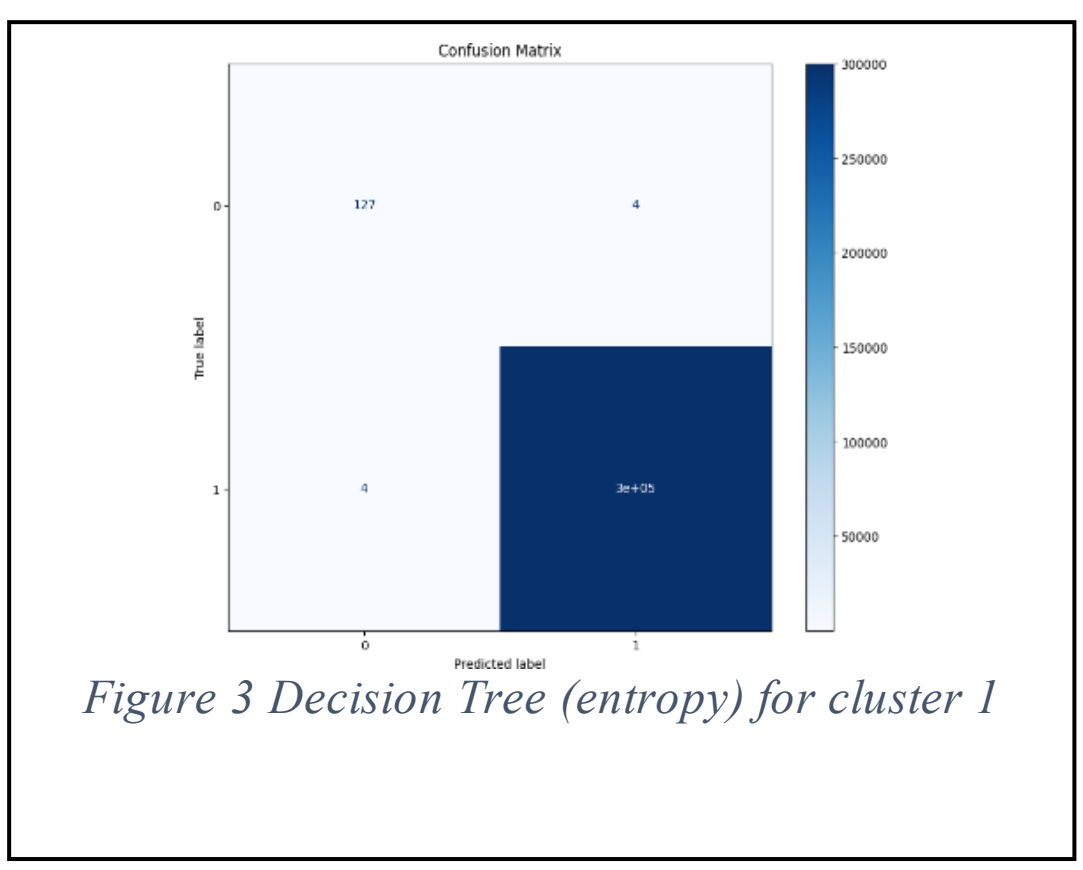

*Figure 3 Decision Tree (entropy) for cluster 1*

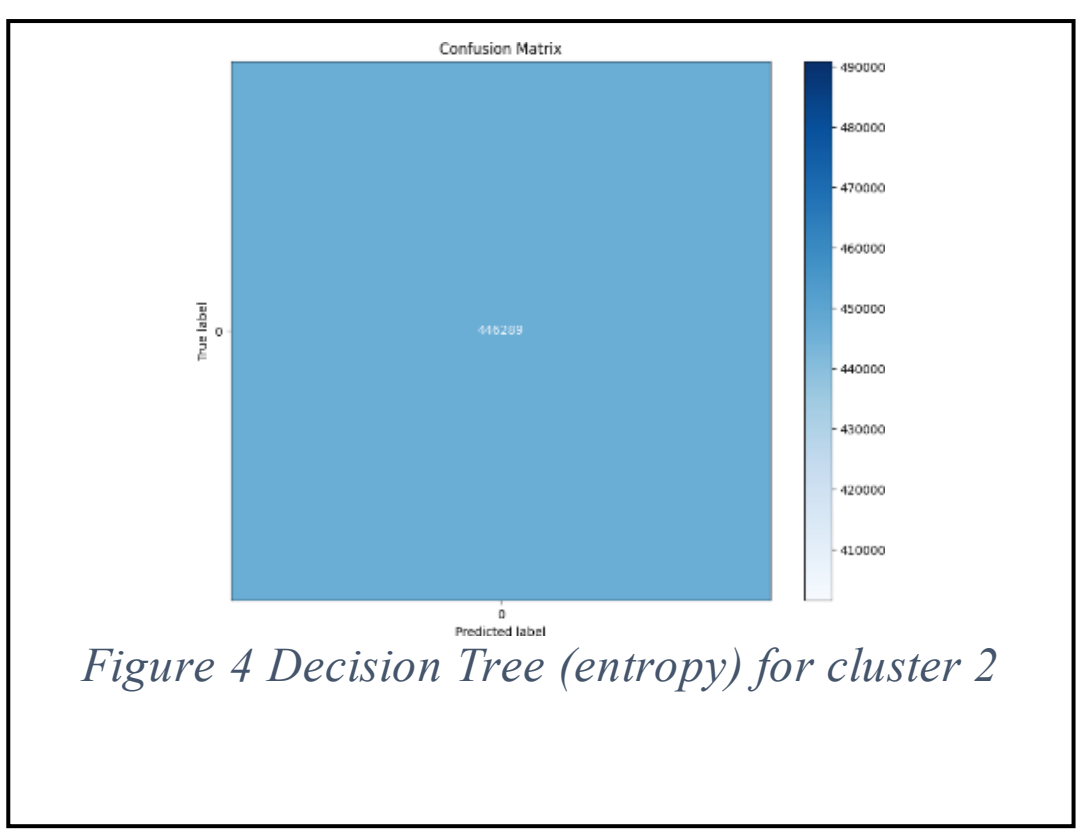

*Figure 4 Decision Tree (entropy) for cluster 2*